# Lowered Evans Models:
# Analytic Distribution Functions of Oblate Halo Potentials

Konrad Kuijken[1,2] and John Dubinski[3]

Harvard–Smithsonian Center for Astrophysics, 60 Garden St., Cambridge, MA 02138

## ABSTRACT

We describe an analytic distribution function of a finite, oblate stellar system that is useful for the practical modelling of dark halos. The function is determined by lowering Evans's (1993) distribution function of a flattened, cored isothermal system in analogy to the lowering of the singular, isothermal sphere in the definition of the King (1966) models. We derive analytic expressions of the density, maximal streaming velocity and velocity dispersion profiles as a function of the potential. As for the King models, the potential must be calculated numerically. We also present a recipe for generating N-body realizations of this distribution function and examine the stability in three models with dimensionless spins $\lambda = 0.0$, 0.05 and 0.18 using N-body simulations with 50,000 particles. The $\lambda = 0.18$ model is unstable to the formation of a triaxial bar within $\sim 5$ King radii while the other models appear stable. We conclude that the slowly rotating systems are useful for modelling flattened dark halos.

*Subject headings:* galaxies: kinematics and dynamics – galaxies: structure – galaxies: Elliptical – methods: numerical

## 1   Introduction

The dark halos surrounding disk galaxies are probably not spherical. In order to simulate the effects of non-spherical halos on disks, it is useful to have an analytic distribution function (DF) for flattened halos. Such models do exist, but they are either not very realistic models for galaxy halos (such as the flattened Plummer models of Lynden-Bell 1962) or infinite in extent (for example, the scale-free logarithmic models of Toomre 1982). In this paper we construct a set of analytic DF's for finite-radius, oblate galactic halo models, and show that they can be sampled efficiently for use as an equilibrium starting condition in N-body simulations.

---

[1] I: kuijken@cfa.harvard.edu

[2] Hubble Fellow

[3] I: dubinski@cfa.harvard.edu

Distribution functions are most useful when specified as functions of integrals of motion. Most realistic galaxy potentials admit three of these, but in the case of axisymmetric models explicit expressions only exist for two (except for the special cases of Stäckel potentials and of spherical models). The DFs that we construct here will therefore be functions only of the two classical integrals of motion, namely the particle energy $E$ and angular momentum about the symmetry axis $L_z$. As shown by Lynden-Bell (1962), there is a unique relation between the even part of the DF, $f(E, |L_z|)$, and the density $\rho(R, \Psi)$ in the meridional plane when expressed as a function of the potential and cylindrical radius. Arbitrary terms odd in $L_z$ may be added to the DF without affecting the density $\rho$: since such terms effectively change the azimuthal direction in which some orbits are traversed, they set the total angular momentum of the system.

In this paper, we modify an existing analytic DF for infinite flattened halo models, due to Evans (1993), making it finite in extent. The resulting models have a range of flattenings, central densities, outer radii and concentration parameters. Evans's model is described in §2, and our modification of it in §3. Section 4 contains a recipe for generating N-body realizations of the models, and the results of N-body experiments to test the stability of several of the models. We find that as long as the mean rotation of the models is kept within reasonable cosmological expectations, there is no sign of large-scale bar instability. A summary is provided in §5.

## 2  Evans's distribution function for the Binney potentials

Evans (1993) found the exact two-integral distribution function (DF) for all axisymmetric Binney (1981) potentials, by applying Lynden-Bell's (1962) Laplace transform method. His models are among the few fully analytic axisymmetric potential-density-DF sets that can be written down, and they are very simple, at that: Binney's potential is

$$\Psi(R, z) = \sigma_0^2 \ln\left[\frac{R_c^2 + R^2 + (z/q)^2}{R_1^2}\right], \quad (1)$$

the mass density is

$$\rho(R, z) = 8\pi G \sigma_0^2 \frac{(3 + q^{-2})R_c^2 + (1 + q^{-2})R^2 + (3q^{-2} - q^{-4})z^2}{(R_c^2 + R^2 + (z/q)^2)^2} \quad (2)$$

and the two-integral DF has the form

$$f(E, L_z) = A L_z^2 \exp(-2E/\sigma_0^2) + B \exp(-2E/\sigma_0^2) + C \exp(-E/\sigma_0^2). \quad (3)$$

The constant $R_1$ (which Evans set to one) is introduced here in the potential for consistency of units. The related density,

$$\rho_1 = \frac{\sigma_0^2}{2\pi G R_1^2} \quad (4)$$

serves as a density scale parameter for the models. The constants $A$, $B$ and $C$ depend on $\rho_1$, the velocity dispersion $\sigma_0$, the axis ratio of the equipotentials $q$ and the core radius $R_c$, as follows:

$$A = \frac{8(1-q^2)G\rho_1^2}{\pi^{1/2}q^2\sigma_0^7}, \qquad B = \frac{4R_c^2 G\rho_1^2}{\pi^{1/2}q^2\sigma_0^5}, \qquad \text{and} \quad C = \frac{(2q^2-1)\rho_1}{(2\pi)^{3/2}q^2\sigma_0^3}. \qquad (5)$$

In particular, spherical models have $A = 0$ and coreless models have $B = 0$. The familiar isothermal sphere DF is recovered when both these constants are zero: in that case, the DF is a Maxwellian with velocity dispersion $\sigma_0$ and density $\rho_1 \exp(-\Psi/\sigma_0^2)$. The coreless, $B = 0$ models were first discovered by Toomre (1982).

## 3 Lowered Evans models

All the models (3) are infinite in extent; moreover, at large radii the mass density falls as inverse-square radius, implying that they have infinite mass. These models are useful for the purposes of modelling galactic halos, since they have an adjustable core radius and a flat rotation curve at large radii. Nevertheless, for many applications it would be more convenient to have a finite-mass model, which nonetheless has a flat rotation curve over an appreciable radius range. In analogy with the lowered isothermal, or King (1966) models, we were therefore motivated to construct finite oblate halo models by 'lowering' Evans's DF, in effect imposing a maximum energy on the stars in the model: thus we write

$$f(E, L_z) = \begin{cases} [(AL_z^2 + B)\exp(-E/\sigma_0^2) + C][\exp(-E/\sigma_0^2) - 1] & \text{if } E < 0, \\ 0 & \text{otherwise.} \end{cases} \qquad (6)$$

Only stars with negative energy are included in these models. There are many other ways of accomplishing an energy cutoff in the DF, or indeed of limiting the DF in a different way (Binney and Tremaine 1987, Kashlinsky 1988, Rowley 1988) but we follow King's lead. The outermost radius at which the potential is negative (and hence the density non-zero) is usually called the *tidal radius*.

The lowered DF (eq. 6) no longer corresponds to a self-gravitating system in Binney's potential (eq. 1). To contruct self-consistent models with this DF, we use the fact that the DF determines the mass density in terms of the gravitational potential:

$$\rho(R, z) = \int dv_R dv_\phi dv_z \; f = \frac{2\pi}{R} \int_{E \geq \Psi + \frac{1}{2}(L_z/R)^2} dL_z dE \; f \equiv \rho(\Psi, R). \qquad (7)$$

Therefore we can achieve self-consistency by combining this expression with the Poisson equation, to yield

$$\nabla^2 \Psi = 4\pi G \rho(\Psi, R), \qquad (8)$$

and solving for the potential.

In the case of the lowered Evans DF, some algebra shows the corresponding density in an arbitrary potential (see eq. 7) to be

$$\begin{aligned}
\rho(\Psi, R) &= \frac{1}{2}\pi^{3/2}\sigma_0^3(AR^2\sigma_0^2 + 2B)\,\mathrm{erf}(\sqrt{-2\Psi}/\sigma_0)\exp(-2\Psi/\sigma_0^2) \\
&\quad + (2\pi)^{3/2}\sigma_0^3(C - B - AR^2\sigma_0^2)\,\mathrm{erf}(\sqrt{-\Psi}/\sigma_0)\exp(-\Psi/\sigma_0^2) \\
&\quad + \pi\sqrt{-2\Psi}[\sigma_0^2(3A\sigma_0^2 R^2 + 2B - 4C) + \frac{4}{3}\Psi(2C - A\sigma_0^2 R^2)],
\end{aligned} \quad (9)$$

where $\mathrm{erf}(x) = 2\pi^{-1/2}\int_0^x \exp(-t^2)\,dt$ is the usual error function. The maximum streaming velocity is obtained by inserting an extra factor of $|L_z|/R$ in the integral (7): then

$$\begin{aligned}
\rho v_{\phi\mathrm{max}} &= 2\pi\Psi^2(AR^2\sigma_0^2 - C) + 2\pi\Psi\sigma_0^2(2C - B - 3AR^2\sigma_0^2) \\
&\quad + \pi\sigma_0^4(7AR^2\sigma_0^2 + 3B - 4C) + 4\pi\sigma_0^4(C - B - 2AR^2\sigma_0^2)\exp(-E/\sigma_0^2) \\
&\quad + \pi\sigma_0^4(AR^2\sigma_0^2 + B)\exp(-2E/\sigma_0^2).
\end{aligned} \quad (10)$$

The velocity dispersions follow similarly. They are given by $\sigma_i^2 = \rho^{-1}\int d^3v\, f v_i^2$:

$$\begin{aligned}
\rho\sigma_R^2 = \rho\sigma_z^2 &= \frac{8}{15}\pi\psi^2\sqrt{-2\psi}[AR^2\sigma_0^2 - 2C] + 2\pi\psi\sigma_0^2\sqrt{-2\psi}[-AR^2\sigma_0^2 - \frac{2}{3}B + \frac{4}{3}C] \\
&\quad + \pi\sigma_0^4\sqrt{-2\psi}[\frac{7}{2}AR^2\sigma_0^2 + 3B - 4C] \\
&\quad + (2\pi)^{3/2}\sigma_0^5[-AR^2\sigma_0^2 - B + C]\,\mathrm{erf}(\sqrt{-\psi}/\sigma_0)\exp(-\psi/\sigma_0^2) \\
&\quad + \frac{1}{4}\pi^{3/2}\sigma_0^5[AR^2\sigma_0^2 + 2B]\,\mathrm{erf}(\sqrt{-2\psi}/\sigma_0)\exp(-2\psi/\sigma_0^2),
\end{aligned} \quad (11)$$

and

$$\begin{aligned}
\rho\sigma_\phi^2 &= \frac{8}{15}\pi\psi^2\sqrt{-2\psi}[3AR^2\sigma_0^2 - 2C] + 2\pi\psi\sigma_0^2\sqrt{-2\psi}[-3AR^2\sigma_0^2 - \frac{2}{3}B + \frac{4}{3}C] \\
&\quad + \pi\sigma_0^4\sqrt{-2\psi}[\frac{21}{2}AR^2\sigma_0^2 + 3B - 4C] \\
&\quad + (2\pi)^{3/2}\sigma_0^5[-3AR^2\sigma_0^2 - B + C]\,\mathrm{erf}(\sqrt{-\psi}/\sigma_0)\exp(-\psi/\sigma_0^2) \\
&\quad + \frac{1}{4}\pi^{3/2}\sigma_0^5[3AR^2\sigma_0^2 + 2B]\,\mathrm{erf}(\sqrt{-2\psi}/\sigma_0)\exp(-2\psi/\sigma_0^2).
\end{aligned} \quad (12)$$

At $R = 0$ all three dispersions are the same (as they are in all non-singular two-integral models). The difference $\rho\sigma_\phi^2 - \rho\sigma_R^2$ indicates the anisotropy of the velocity dispersion. This expression is proportional to $AR^2\sigma_0^2$ (the case $A = 0$ corresponds to spherical models, since the $B$- and $C$-terms are isotropic), with a coefficient that is positive for all $\Psi < 0$: therefore all oblate non-rotating models ($A > 0$ in eq. 3) satisfy $\sigma_\phi \geq \sigma_R$ everywhere, while the converse holds for the prolate models.

The King (1966) models are just the special case $A = B = 0$. These models are spherical, with $R_c = 0$. (Note the distinction between the core radius $R_c$ of the Binney potential before lowering the DF, and the 'King' radius of the density, which measures the core radius of the self-consistent mass density of the model. Both these radii lose their meaning somewhat in the lowered models.) In the spherical case it is possible to solve Poisson's equation (eq. 8) as a simple forward integration in radius. Given choices for the density scale $\rho_1$ and the velocity dispersion parameter $\sigma_0$, the central potential energy $\Psi_0$ (we have fixed the cutoff energy at zero) may be specified as a boundary condition for this integration. Deeper central potential depth implies a higher central density and a larger tidal radius, and therefore leads to a more centrally condensed model.

Calculating the two-dimensional models is a little harder, since we cannot now integrate the Poisson equation directly. However, the principle remains the same: once we have chosen the scale parameters $\sigma_0$ and $R_c$ and the flattening $q$, every value of the central potential (or density) implies a unique non-singular self-consistent model. In practice, numerical iterations are required to obtain the density and potential: a guess is made at the potential, the density corresponding to the DF in this potential is calculated from eq. 6, Poisson's equation is solved for the potential corresponding to this density, and this new potential is taken as the start of the next iteration. In this work we have used a multipole expansion (Prendergast & Tomer 1970) to solve for the potential: the models are quite smooth, and hence it was profitable to solve for the radial dependencies of the dominant harmonic terms (a few one-dimensional functions) rather than attacking a two-dimensional grid calculation. For finite-extent systems the multipole expansion has the further advantage that boundary conditions at infinity are automatically handled correctly. To ease numerical convergence, the higher harmonics were introduced one at a time, allowing a few iterations for any oscillations to stabilize before the next term was added. Potentials of the models were calculated up to $l = 4$, and the densities were then derived using eq. (9).

King models have non-singular cores, in spite of the fact that the DF does not contain the $B$-term (the only term proportional to the Binney potential core radius $R_c$). For very deep central potentials these models have sharply peaked densities, though, asymptoting to the singular isothermal sphere in their central regions. The central core radius, or King radius, of the King models' potential is related to the central density $\rho_c$ by

$$r_K = \left( \frac{9\sigma_0^2}{4\pi G \rho_c} \right)^{1/2}. \tag{13}$$

At this radius the gravitational potential has risen by about $2\sigma_0^2$ over the central value, provided the potential well depth is well below $\sigma_0^2$. A concentration parameter is usually defined as the ratio of the outer radius of the model and the King radius.

The flattened models with $B = 0$ also become more centrally concentrated as the depth of the potential is increased. However, these models are not very satisfactory: the central density contours have undesirable depressions on the symmetry axis. It turns out that a small $B$-term (i.e., non-zero $R_c$) suffices to remove this effect, since the addition of this isotropic term, more centrally concentrated than the $C$-term, serves to round off the central density figure. Some experimentation shows that the choice $R_c = r_K$ results in a more elliptical central core, but does not affect the flattening of the halo model at larger energies too drastically. Addition of a $B$-term affects the central density: in practice, a quadratic equation in $R_c^2$ needs to be solved to find the core radius that yields a central density whose

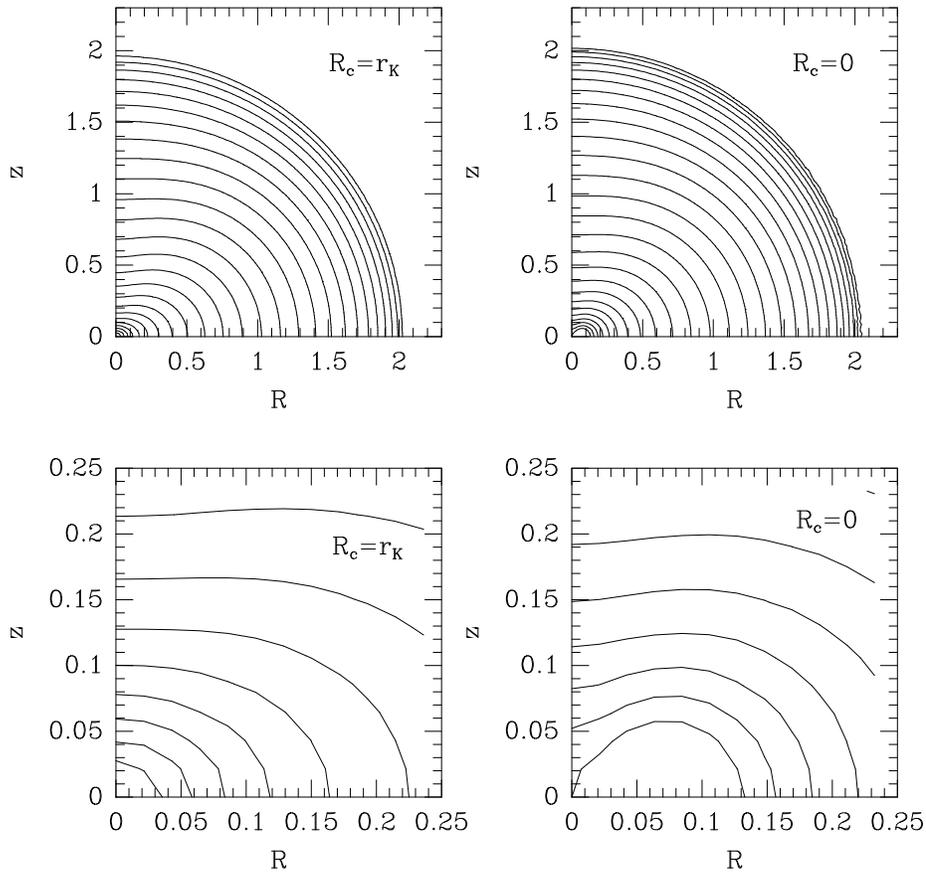

Fig. 1.—Iso-density contours for two $q = 0.8$, $\Psi_0 = -6\sigma_0^2$ models. The density on adjacent contours differs by a factor of two. The left-hand panels show the model with $B$-term (which has King radius 0.053), the right-hand panels the one without (in this case $r_K = 0.163$). Both models have similar shape outside the core, but the addition of the $B$-term removes the strong 'peanut' dimples in the central regions.

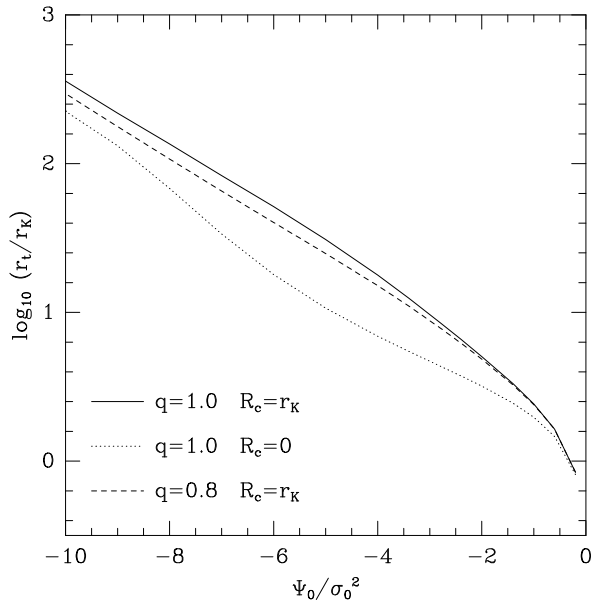

FIG. 2.–Concentration vs. central potential, for different choices of flattening. The dotted line represents the original King (1966) models.

King radius is equal to $R_c$. The isodensity contours for $q = 0.8$, $\Psi_0 = -6\sigma_0^2$ models with and without such a $B$-term are shown in Figure 1, illustrating the rounding effect in the central regions. For shallower models ($\Psi_0 \gtrsim -3\sigma_0^2$), it is better to take $R_c = kr_K$ with $k < 1$, otherwise all the models end up rather round.

Our models, in summary, are the self-consistent realizations of the axisymmetric distribution functions (6). The parameters $\sigma_0$, $\rho_1$, $\Psi_0$ and $q$ may be chosen freely, after which the constant $R_c$ is adjusted until the central density $\rho_c$ is equal to $(9k^2\sigma_0^2/4\pi G R_c^2)$, i.e. until the King radius is equal to $R_c/k$, where $k$ lies between 0.3 and 1.

The choice of the four parameters $\rho_1$, $\sigma_0$, $q$ and $\Psi_0$ is equivalent to picking a tidal radius, a concentration parameter, a flattening and a central density (or a total mass) for the system. Figures 2 and 3 present some of the relevant relations: they show the dependence of the concentration parameter $r_t/r_K$ and of the scaled central density as a function of the scaled potential well depth, $\Psi_0/\sigma_0^2$.

Figure 4 shows the circular velocity curve in the equatorial plane of the models of Figure 1.

## 4  N-body Realizations

The lowered Evans model is useful for examining galactic dynamical problems that depend on flattened potentials such as disk warping or polar rings. The model is finite in extent making it ideal for use in N-body simulations. In this section, we present a recipe for generating an N-body realization of a lowered Evans model. We then test the stability of three sample models using $q = 0.8$ with different amounts of spin: a non-rotating model, a maximally streaming model and a model with spin corresponding to the cosmological expectations of the dimensionless spin parameter, $\lambda = 0.05$.

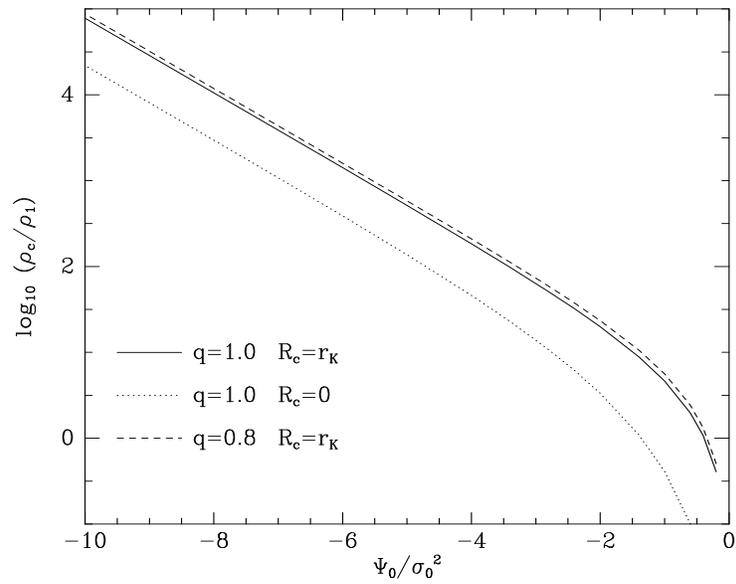

FIG. 3.–Central density vs. central potential for the models indicated.

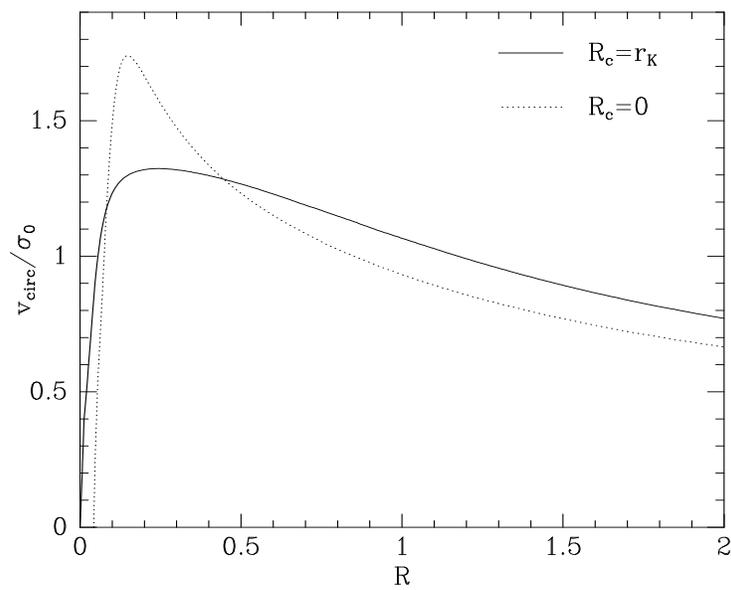

FIG. 4.–Circular velocity curves in the equatorial plane of the $q = 0.8$, $\Psi_0 = -6\sigma_0^2$ models.

### 4.1 Initial Conditions

We can generate an N-body realization from any DF of the form $f(E, L_z)$ by sampling from it in two stages. First, we sample values of $R$ and $z$ from the density distribution, $\rho(R, z)$ to find the particle positions. Then, for each position the DF and the gravitational potential define the distribution of velocities, and we sample from this function to assign each particle's velocity.

We use the acceptance-rejection technique for sampling the distributions (e.g., Press et al. 1993), which works as follows. Let $F_{\max}$ be the maximum value of the distribution function. Then, we randomly select, with a uniform distribution, a point in the allowed domain of the independent variable(s), and we also sample a 'test' value $F$ from the range $[0, F_{\max}]$ of the distribution function. The value of $F$ determines whether the sampled point will be included or not: if $F$ is less than the value of the distribution function at the sampled point, that point is accepted as valid, otherwise it is rejected and another point is sampled. If the distribution being sampled is very non-uniform, this process may be quite inefficient, with many of the sampled points being rejected. To avoid this situation, it is worthwhile to transform the independent variables to coordinates whose distribution is as uniform as possible.

The distribution of the particle positions (up to a normalization) is the density $\rho(R, z)$. For the lowered Evans model, $\rho \sim r^{-2}$, so it is convenient to introduce the variable $u = \tan^{-1} z/R$ which makes the mass element almost uniform:

$$\rho(R, z)dRdz = \rho(u, R)(R^2 + z^2)dRdu. \tag{14}$$

The domains of $R$ and $u$ are $0 \leq R \leq r_t$ and $-\pi/2 < u < \pi/2$. For every $(u, R)$ point sampled, we also sample an independent random azimuthal angle $\phi$, finally allowing us to define $x = R\cos\phi$, $y = R\sin\phi$ and $z = R\tan u$. Each particle thus sampled is then assigned a velocity vector, in essentially the same way: the velocity distribution function at each particle's position is known from the DF (eq. 6), where $E = \frac{1}{2}v^2 + \Psi(R, z)$ and $L = Rv_\phi$, and the velocity is always less than the local escape speed, $v_{\rm esc} = \sqrt{-2\Psi}$. The velocity vector can thus be sampled with the acceptance-rejection technique from inside a sphere of this radius.

The DF of eq. (6) does not depend on the sign of $L_z$: consequently it has no net streaming. We can introduce azimuthal streaming into the model by varying the number of particles with $v_\phi$ positive and negative. In a non-rotating model, there are equal numbers of particles with $v_\phi$ going in opposite directions while for a maximally streaming model, the sign of $v_\phi$ is the same for all the particles. All intermediate cases have varying fractions of $v_\phi$ going in opposite directions.

### 4.2 Stability

Although it is easy to formulate an equilibrium DF, it is not guaranteed that the resulting system is stable. We therefore tested the stability of rotating and non-rotating models with 50,000-particle N-body simulations. Initial conditions were generated with the procedure described above (Figure 5). We worked in units in which $G = 1$ and $\rho_1 = (4\pi)^{-1}$, and investigated models with $\sigma_0 = 2^{-1/2}$, flattening $q = 0.8$ and central potential $\Psi_0 = -6.0\sigma_0^2$. The parameter $R_c$ was chosen to be equal to the King radius, as described in §3. The base model has the following properties: (i) The central density is $\rho_c = 126$ so that the equivalent King radius from equation (13) is $r_K = 0.053$, (ii) the model extends to a "tidal" radius $r_t = 2.14$, (iii) the core crossing time is $T_{core} = (3\pi/G\rho_c)^{1/2} = 0.30$, and (iv) the system crossing time is $T_{sys} = 2r_t/\sigma_0 = 6.0$. We introduce varying degrees of rotation in three models by varying the fraction of particles going in opposite directions. We parameterize the rotation using the dimensionless spin parameter, $\lambda = G^{-1}L|E|^{1/2}M^{-5/2}$, that is used to quantify spin in dark halos formed in cosmological models. The typical value for a cosmological dark halo is thought to be $\lambda = 0.05$ (e.g. Barnes & Efstathiou 1987; Warren et al. 1992). The three models we investigated have $\lambda = 0.0$ (a non-rotating halo), $\lambda = 0.05$ (a "cosmological" halo) and $\lambda = 0.18$ (the maximally streaming halo).

We then used a tree code (Barnes & Hut 1986; Dubinski 1988) to simulate the models for 24 units of time corresponding to 80 core crossing times and 4 system crossing times. We used an opening angle tolerance, $\theta = 1.0$ and calculate cell-particle forces to quadrupole order with a particle softening radius, $r_{soft} = 0.005$. We integrate the trajectories using a leapfrog integrator with a timestep $\Delta t = 0.02$ corresponding to 12 steps per core crossing time. The error in the total energy of the systems was no more than 2.5% by then end of each run.

We measured the stability of the models by comparing the density and velocity dispersion profiles, averaged over spherical shells, at early and late times. Shell averages are trivial to compute from N-body simulations, requiring a simple binning of the particles by radius. The exact shell averages that correspond to the initial analytic DF are also straightforward to obtain: one can show that for a given cylindrically symmetric function $g(R, z)$ the spherically averaged function $\overline{g}(r)$ found by averaging the function $g$ within a thin spherical shell at radius $r$ is

$$\overline{g}(r) = \frac{1}{r}\int_0^r g([r^2 - z^2]^{1/2}, z)dz. \quad (15)$$

We can therefore calculate the averaged density profile $\overline{\rho}(r)$ and velocity dispersion profiles $\overline{\sigma_R^2}$, $\overline{\sigma_z^2}$, and $\overline{\sigma_\phi^2}$ by inserting the functions from equations (9) and (11), 12) into this integral.

Figure 6 compares the expected averaged density profile to measurements from the rotating and non-rotating models at the final time $t = 24.0$. The density profiles remain unchanged over a large range in radii except near the center of the models. The density declines slightly in the center, probably in response to the greater degree of integration error within the core. Nevertheless, the mass profiles of the models appear stable over at least four system crossing times.

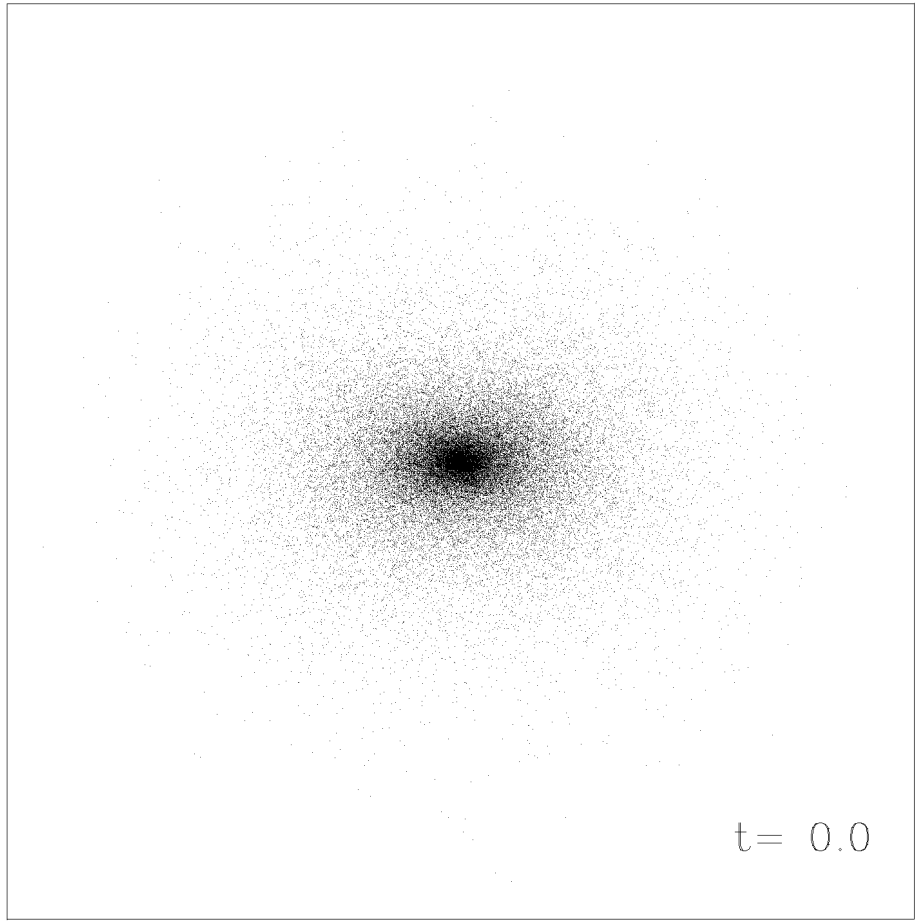

Fig. 5.—Edge on view of the 50,000 particle model used in the simulations. The box width is 4.0 units across (80 King radii). The flattening in the density is about $q_\rho = 0.6$ in the center increasing to $q_\rho = 0.8$ at the tidal radius.

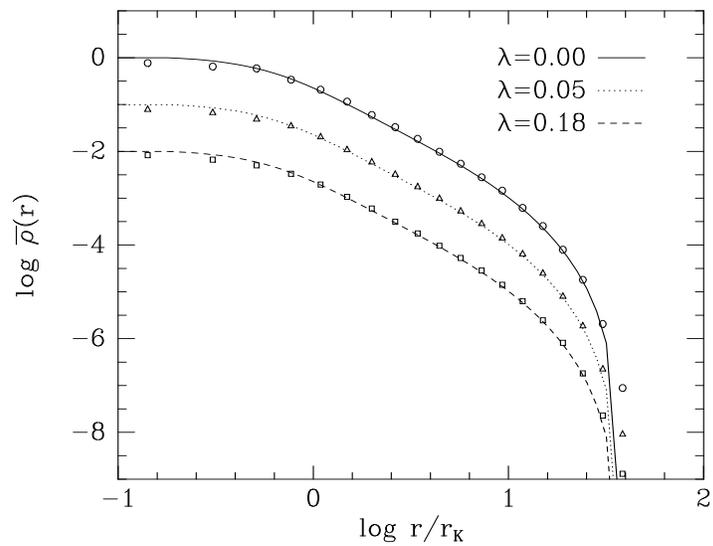

Fig. 6.—The spherically averaged density profiles for the 3 models normalized to the central density. The profiles of the different models are offset by one dex for illustration. The curves are the theoretical expectation and the points are the density at the end of the simulation, estimated by binning particles in spherical shells. The agreement is very good over a large range in radii except for a slight dip in the core that is probably due to integration error.

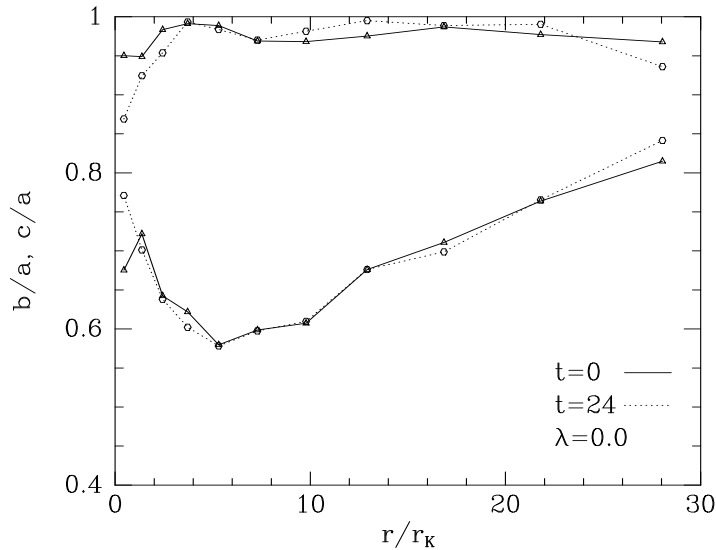

Fig. 7a.–Axial ratio profiles of density contours at $t = 0.0$ and $t = 24.0$ for the three models. The profiles for the $\lambda = 0.0$ and $\lambda = 0.05$ models remain fairly constant. There is a hint of a bar forming within $r = r_K$ though it is difficult to judge if this is real because of integration and measurement error. The $\lambda = 0.18$ model shows a distinct bar out to $r \simeq 5r_K$ indicating the onset of the bar instability.

A more refined indication of stability is provided by a study of quadrupole terms in the mass distribution. We therefore estimate the axial ratio profile of the dark halos using the technique described in Dubinski & Carlberg (1991). In this algorithm, initial values for the axis ratios $q_1$ and $q_2$ are assumed, and used to calculate a starting approximation to the modified inertia tensor, $I_{ij} = \sum x_i x_j / a^2$ for particles in ellipsoidal shells of axial ratios $q_1$ and $q_2$ ($x_i$ is the particle position and $a^2 = x^2 + y^2/q_1^2 + z^2/q_2^2$ is the particle elliptical radius). New axial ratios, and the orientation of the ellipsoidal figure, are then estimated from $I_{ij}$ through $q_1^2 = I_{yy}/I_{xx}$ and $q_2^2 = I_{zz}/I_{xx}$, and used to calculate an improved approximation to the modified inertia tensor. Starting with particles in a spherical shell ($q_1 = q_2 = 1$), this process is repeated for several iterations until the axial ratios and the shell orientation converge to values within a specified tolerance ($\Delta q = .001$).

Figure 7 presents the axial ratio profiles measured from the simulations at the initial and final times. The isodensity contours of the dark halo are slightly peanut shaped near the center so that the estimate of the axial ratio from the modified inertia tensor (which assumes that the density contours are ellipsoidal) will underestimate the ratio of the extent of the isodensity contours along the $R$ and $z$ axes somewhat. The axial ratio profile does not change dramatically for the $\lambda = 0.0$ and $\lambda = 0.05$ models suggesting that these systems are not susceptible to strong bar instabilities, though a careful look at the $\lambda = 0.0$ and $\lambda = 0.05$ halos reveal a small hint of a bar within one King radius. Since it did not extend far beyond the core, it is difficult to say whether this feature is real, or a result of integration error and particle noise. On the other hand, the $\lambda = 0.18$ model is clearly unstable, forming a distinct triaxial central bar with an axis ratio $q_1 \simeq 0.7$ and $q_2 \simeq 0.5$ within 5 King radii. A particle plot at the end of the simulation (Figure 8) show this bar clearly: inspection of earlier snapshots reveals that the bar grows during the first system crossing time. It is not too surprising that the $\lambda = 0.18$ model is bar-unstable, since this type of instability generically affects fast-rotating systems: though it is perhaps remarkable that even a system as hot as this one cannot quench it.

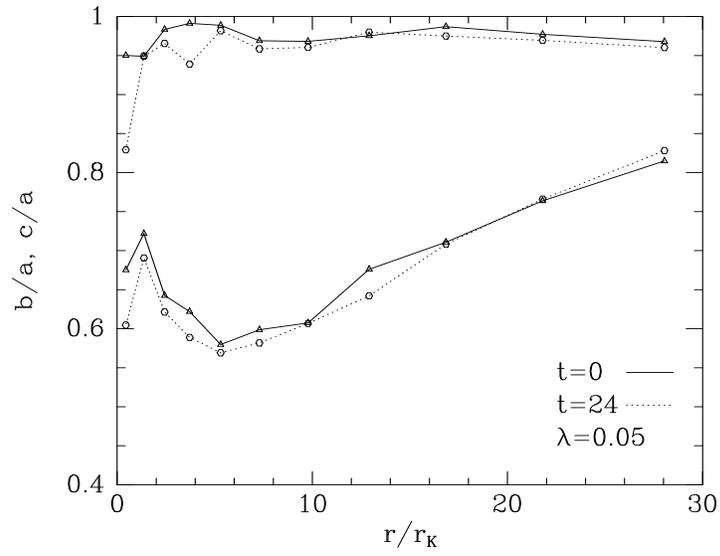

Fig. 7b.—

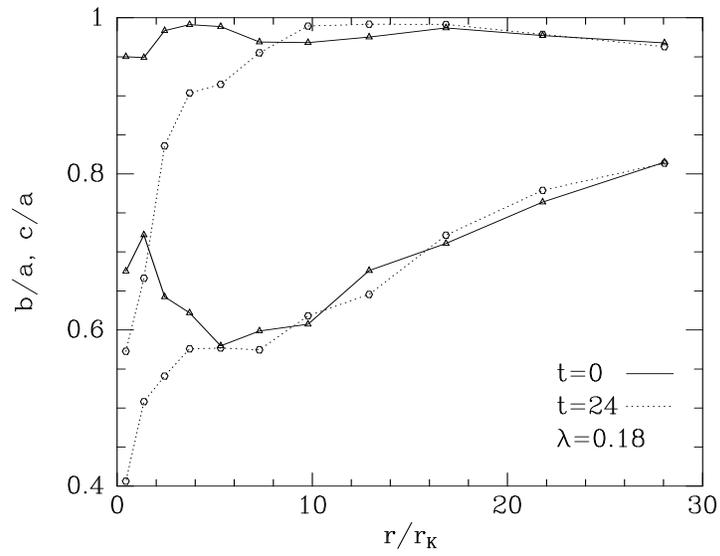

Fig. 7c.—

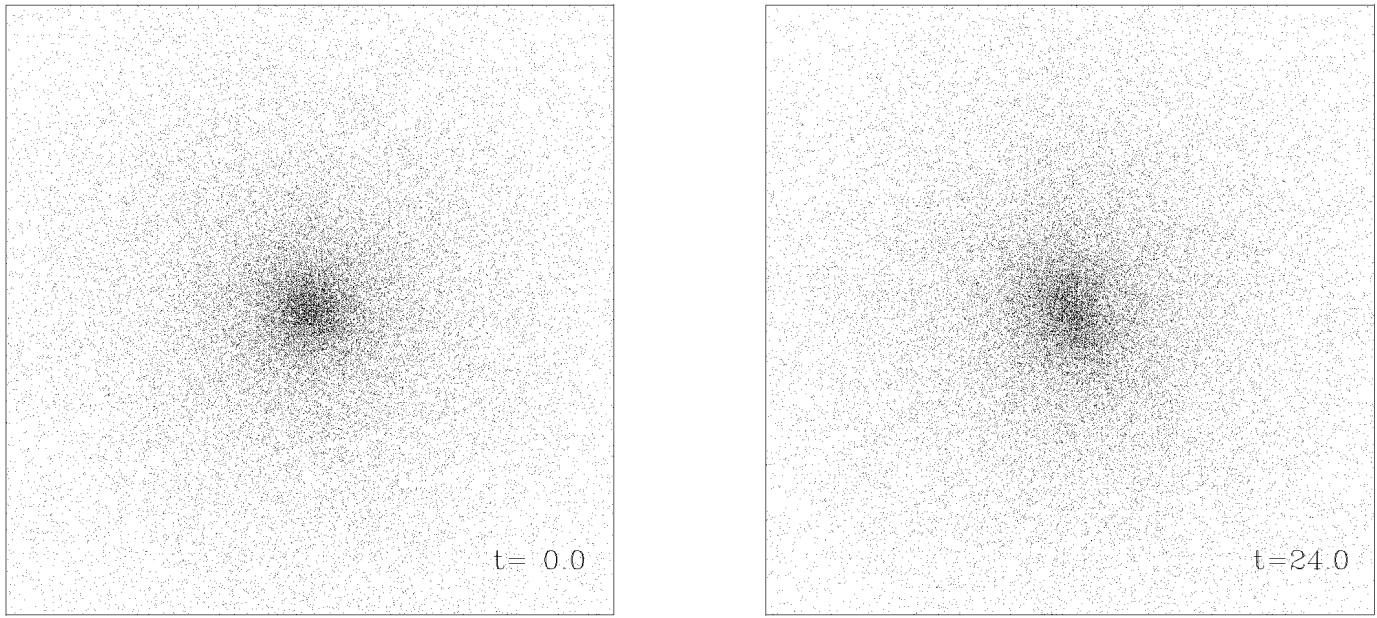

FIG. 8.—Particle plot of the center of the $\lambda = 0.18$ as viewed down the $z$-axis at the $t = 0.0$ and $t = 24.0$. The length of the box is $l = 20r_k$. At $t = 24.0$ a triaxial bar is visible in the center of the system.

The spherically-averaged velocity dispersion profiles (Figure 9) provide further evidence for the stability of the slowly-rotating models, though there is a slight deviation within a King radius, again probably reflecting some integration error within the core. The instability in the $\lambda = 0.18$ model, in contrast, shows up clearly as it develops a noticeable dip in the $\sigma_z$ profile.

In conclusion, the lowered Evans models appear to be stable when the rotation rate is small, but they may suffer bar instabilities in the extreme case of maximal streaming. We recommend the non-rotating and cosmological halos for application to halo modelling but urge caution in the use of more rapidly rotating models. In any case, the bar instability becomes apparent within a crossing time and can be checked in practice before applying the model to a problem.

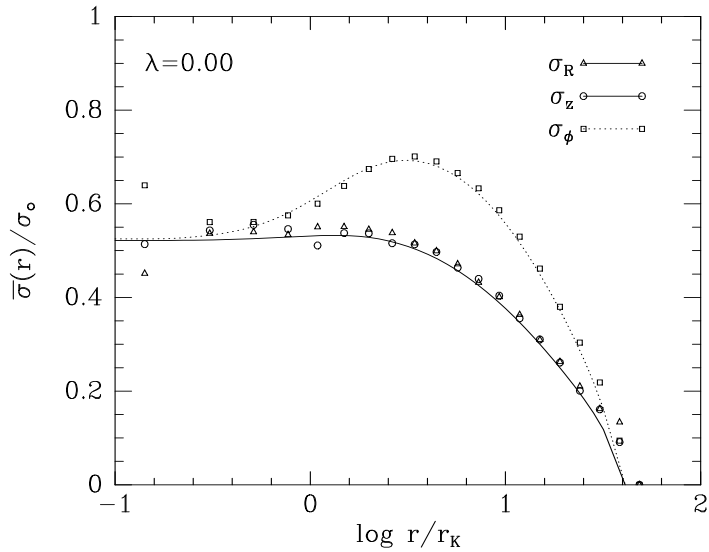

Fig. 9a.—Spherically averaged velocity dispersion profiles for the three models. The curves represent the theoretical shell averaged profiles while the points are the dispersion of the $R, \phi$ and $z$ components of velocity at the end of the simulation, measured by binning particles in shells. The profiles remain unchanged for the $\lambda = 0.0$ and $\lambda = 0.05$ models over most radii though there is a slight deviation within the core again a reflection of integration and measurement error. The systematic dip in $\sigma_z$ in the $\lambda = 0.18$ model is another manifestation of the bar instability.

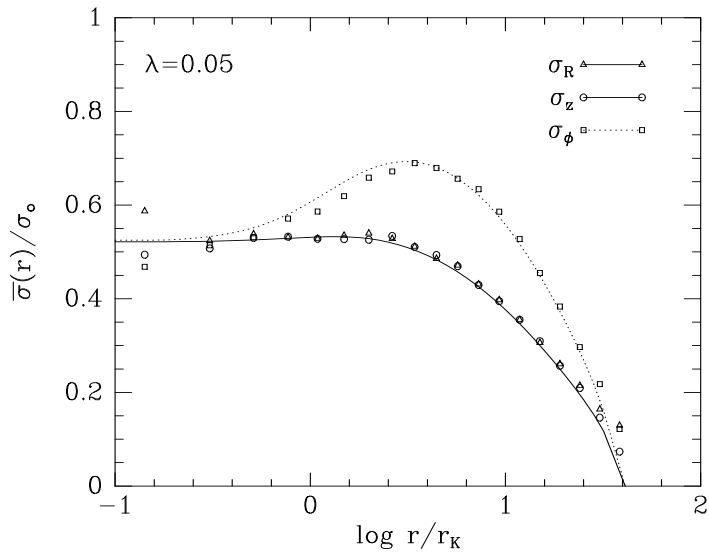

Fig. 9b.–

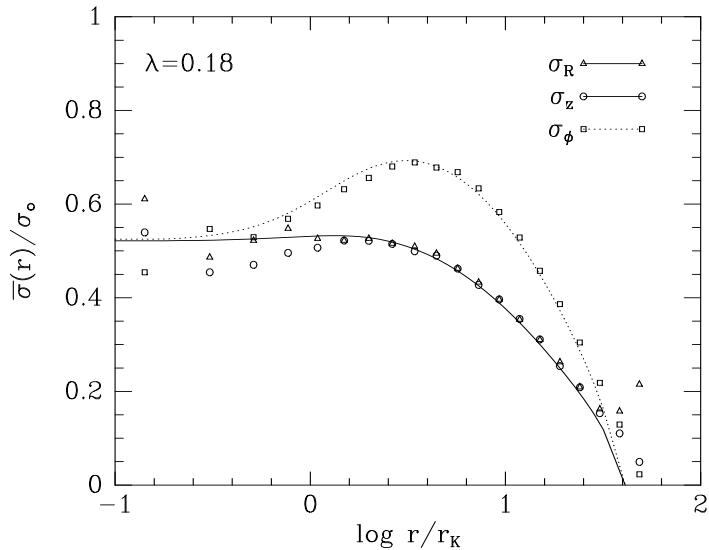

Fig. 9c.–

## 5 Summary

The lowered Evans distribution function provides equilibrium oblate stellar systems which are a flattened analogue of the spherical King models. The density and velocity dispersion profiles can be expressed analytically in terms of the potential though the calculation of the potential still requires a fair amount of numerical computational effort. The payoff for this effort is a finite model ideal for N-body simulations of galactic systems involving flattened dark halos. Furthermore, we provide a simple recipe for generating an N-body realization of the distribution function. A sample of N-body simulations shows that the models are stable for slowly rotating models with spin corresponding to cosmological dark halos. The maximally streaming model is unstable to bar formation despite its high dynamical temperature. We therefore caution users of these models to watch out for the bar instability in rapidly rotating models. In the near future, we plan to apply these models to the formation of warps in disk/halo systems.

### Acknowledgments

We thank Eyal Maoz for comments on the manuscript. JD acknowledges the support of a CfA Postdoctoral Fellowship and KK acknowledges the support of a Hubble Fellowship through grant HF-1020.01-91A awarded by the Space Telescope Science Institute (which is operated by the Association of Universities for Research in Astronomy, Inc., for NASA under contract NAS5-26555).

## REFERENCES

Barnes, J. & Hut, P. 1986, Nature 324, 446
Barnes, J. & Efstathiou, G. 1987, ApJ 319, 575
Binney, J.J. 1981, MNRAS 196, 455
Binney, J. and Tremaine, S. 1987. Galactic Dynamics (Princeton University Press: Princeton), §4
Dubinski, J., 1988, M.Sc. Thesis, University of Toronto
Dubinski, J., & Carlberg, R. G. 1991, ApJ 378, 496
Evans, N.W. 1993, MNRAS 260, 191
Kashlinsky, A. 1988, ApJ 325, 566
King, I.R. 1966, AJ 67, 471
Lynden-Bell, D. 1962, MNRAS 123, 447
Prendergast, K.H. & Tomer, E. 1970, AJ 75, 674
Press, W. H., Teukolsky, S. A., Vetterling, W. T., & Flannery, B. P. 1993, Numerical Recipes in C (Cambridge Univ. Press: Cambridge), §7
Rowley, G. 1988, ApJ 331, 124
Toomre, A. 1982, ApJ 259, 535
Warren, M.S., Quinn, P. J., Salmon, J. K., & Zurek, W. H. 1992, ApJ 399, 405